\documentclass[preprint,12pt,3p]{elsarticle}
\usepackage[utf8]{inputenc} % allow utf-8 input
\usepackage[T1]{fontenc}    % use 8-bit T1 fonts
\usepackage{hyperref}       % hyperlinks
\usepackage{url}            % simple URL typesetting
\usepackage{booktabs}       % professional-quality tables
\usepackage{amsfonts}       % blackboard math symbols
\usepackage{nicefrac}       % compact symbols for 1/2, etc.
\usepackage{microtype}      % microtypography
\usepackage{lipsum}
\usepackage{amsmath,amssymb,graphicx}
\usepackage{float}
\usepackage{siunitx}
\usepackage{array}
\usepackage{lmodern}
\usepackage{amssymb}
\usepackage{xcolor}
\usepackage{multicol}
\usepackage{multirow}
\journal{Elsevier}
\begin{document}
\begin{frontmatter}
\title{Dynamical Formation of Graphene and Graphane Nanoscrolls.}
\author[FTUnB]{M. L. Pereira J\'unior\corref{author}}
\ead{marcelo.lopes@unb.br}
\cortext[author]{I am corresponding author}
\address[FTUnB]{Department of Electrical Engineering, Faculty of Technology, University of Bras\'{i}lia, 70910-900, Bras\'{i}lia, Brazil.}
\author[UnB]{L. A. Ribeiro J\'unior\corref{author}}
\ead{ribeirojr@unb.br}
\address[UnB]{Institute of Physics, University of Bras\'ilia, 70910-900, Bras\'ilia, Brazil}
\author[UNICAMP1,UNICAMP2]{D. S. Galv\~ao}
\address[UNICAMP1]{Applied Physics Department, University of Campinas, Campinas, S\~ao Paulo, Brazil}
\address[UNICAMP2]{Center for Computing in Engineering and Sciences, University of Campinas, Campinas, S\~ao Paulo, Brazil.}
\author[IFPI,UNICAMP1]{J. M. De Sousa\corref{author}}
\ead{josemoreiradesousa@ifpi.edu.br}
\cortext[author]{I am corresponding author}
\address[IFPI]{Instituto Federal de Educação, Ciência e Tecnologia do Piauí - IFPI, Primavera, S\~ao Raimundo Nonato, Piau\'i, 64770-000, Brazil}
\address[UNICAMP]{Applied Physics Department, State University of Campinas -- UNICAMP,  Campinas, 13083-859, SP, Brazil.} 

\begin{abstract}
Carbon nanoscrolls (CNSs) are nanomaterials with geometry resembling graphene layers rolled up into a spiral (papyrus-like) form. Effects of hydrogenation and temperature on the self-scrolling process of two nanoribbons interacting with a carbon nanotube (CNT) have been studied by molecular dynamics simulations for three configurations: (1) graphene/graphene/CNT; (2) graphene/graphane/CNT, and (3) graphane/graphane/CNT. Graphane refers to a fully hydrogenated graphene nanoribbon. Nanoscroll formation is observed for configurations (1) and (2) for temperatures 300-1000 K, while nanoribbons wrap CNT without nanoscroll formation for configuration (3).
\end{abstract}

\begin{keyword}
%% keywords here, in the form: keyword \sep keyword
Reactive Molecular Dynamics \sep Graphene Nanoribbons \sep Graphane Nanoribbons  \sep Carbon Nanoscrolls \sep Self-Scrolling
\end{keyword}

\end{frontmatter}
\section{Introduction}
\label{INT}

After isolation of a single layer of graphite (Graphene) \cite{novoselov2004electric}, obtained by mechanical exfoliation, there is a renewed interest in obtaining new 2D materials \cite{zhang2005experimental,balandin2008superior}. From a topological point of view, graphene layers can be considered as the basis to generate other structures, such as carbon nanotubes \cite{iijima1993single} and nanoscrolls \cite{bacon1960growth}. Carbon nanotubes are understood as graphene layers rolled up into cylinders, while nanoscrolls are graphene layers rolled up into a papyrus-like form. There are many theoretical and experimental works investigating nanoscrolls of different materials \cite{tomanek2002mesoscopic,lisa2003viculis,viculis2003chemical,braga2004structure,coluci2007prediction,shi2010tunable,uhm2020structural,perim2013controlled,junior2021reactive,tromer_PE,perim_Frontiers}.

In contrast to carbon nanotubes (CNTs), carbon nanoscrolls (CNSs) have open ends, which allows an easily radial expansion (tunable interlayer distances) explored in different applications \cite{zhang2010carbon,shi2010mechanics}. The most significant barrier in their usage has been the limited synthesis of high-quality structures. It leads to a controlling inability of some crucial aspects, among them, the number of scrolled layers \cite{liu2016layered,savin2016symmetric,savin2017graphene}. To overcome this barrier, new alternative routes in obtaining CNSs that allow higher control over the final structure to broaden their applications are desirable \cite{kuroda2011one,tojo2013controlled,shin2014ice}. CNSs exhibit distinct electronic and mechanical behaviors when compared with graphene and CNTs \cite{chen2007structural,song2013atomic}. The CNSs nucleus presents vibrational properties that can be useful in developing nano-actuators and electromechanical devices with excellent performance as cathodes in lithium-sulfur batteries \cite{shi2009gigahertz,shi2010translational,zuo2016novel}. It is worth mentioning that Density Functional Theory (DFT) calculations were used to address the synthetic growth concept of carbon-based materials \cite{freitas_JPCC}.

Recently, molecular dynamics simulations have been used to investigate other possible synthetic routes \cite{zhang2010carbon,perim2013controlled,zhang2016formation}. Results showed that CNT triggers the self-scrolling process of graphene nanoribbons. As a result, the layer spontaneously rolls itself around the CNT to lower the surface energy, a process governed by van der Waals interactions. The advantage of this method is that it imposes dry, non-chemical, and room-temperature conditions \cite{zhang2010carbon,perim2013controlled,zhang2016formation}. An experimental study was conducted by \textit{Xie et. al.} to obtain stable CNSs focused on offering higher control over the final product \cite{xie2009controlled}. In their method, graphite is mechanically exfoliated, deposited over SiO$_2$ substrate, and then immersed in a solution of water and isopropyl alcohol. After resting for a few minutes, the system spontaneously formed CNSs \cite{xie2009controlled}. Another possibility is to use functionalized structures such as graphene oxides and hydrogenated ones \cite{zuo2016novel,amadei_nanoscale}.

Herein, we used fully-atomistic reactive (ReaxFF) molecular dynamics (MD) simulations to investigate the effects of hydrogenation on the self-scrolling mechanism of graphene and graphane nanoribbons when triggered by CNTs. Our results show that it is possible to obtain stable CNT-wrapped CNSs from pure and hydrogenated graphane nanoribbons.

\section{Computational Methodology}
\label{CM}

\begin{figure*}[h!]
\centering
\includegraphics[width=0.8\linewidth]{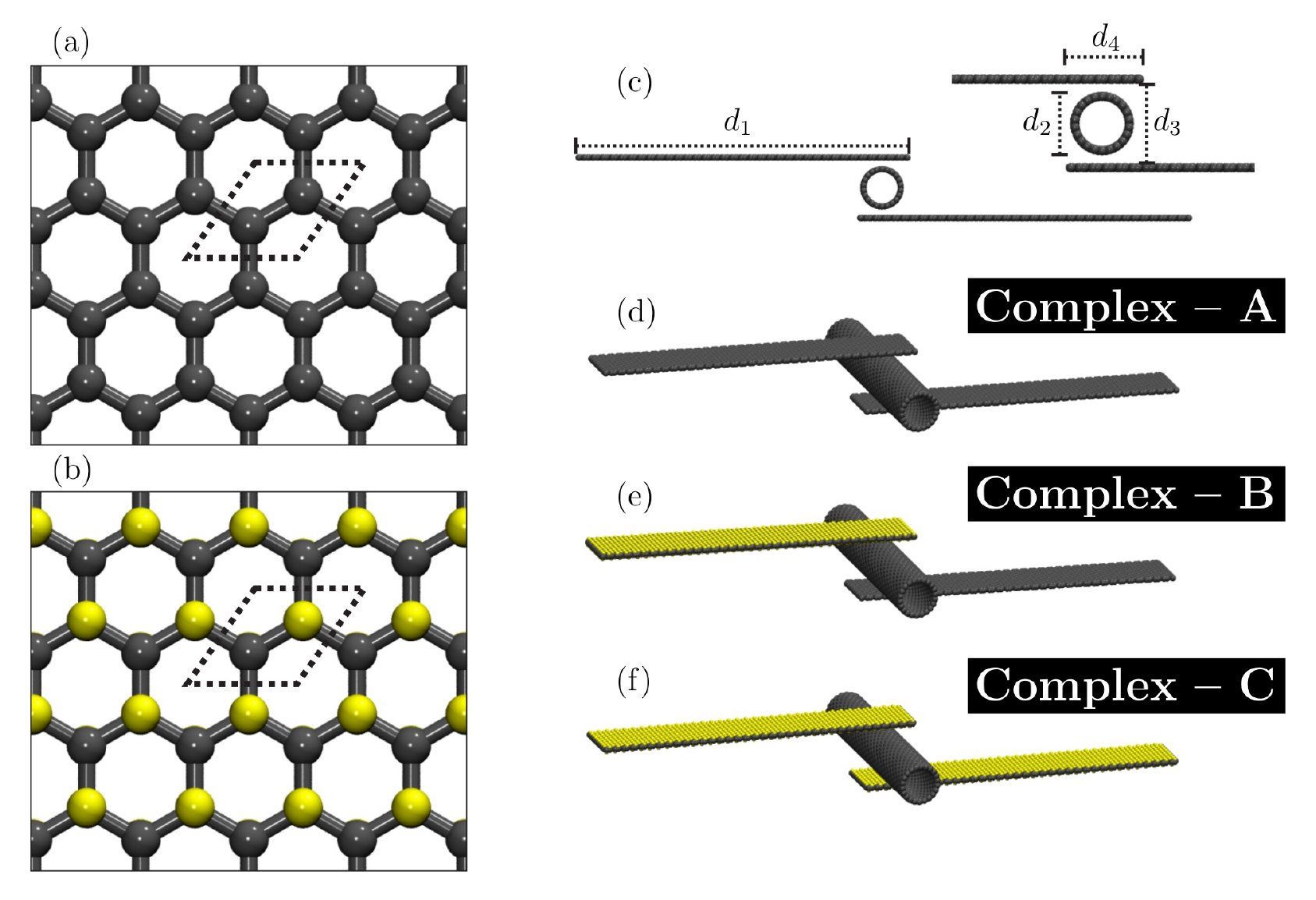}
\caption{Model systems used to simulate the formation of CNT-wrapped CNSs. Illustration of the atomic configuration for (a) graphene and (b) graphane nanoribbons. The dashed lines highlight the structural units. (c) Diagrammatic representation for the initial arrangement of the system involving two nanoribbons and a CNT, where $d_1=150$ \AA~(nanoribbon length), $d_2=17.21$ \r{A} (CNT diameter), $d_3=28.03$ \r{A} (initial distance between the nanoribbons), and $d_4=23.00$ \r{A} (ribbon length into contact with nanotube). Initial configurations for the systems studied here: (d) graphene/graphene/CNT (Complex--A), (e) graphene/graphane/CNT (Complex--B), (f) and graphane/graphane/CNT (Complex--C). In the color scheme, the gray and yellow spheres represent carbon and hydrogen atoms, respectively.}
\label{fig:complexes}
\end{figure*}

In the present work, we used fully atomistic (ReaxFF) molecular dynamics simulations for C/H/O \cite{ashraf2017extension}, as implemented in LAMMPS code \cite{plimpton1995fast}. The initial configurations of our model systems are presented in Figure \ref{fig:complexes}, where we shown three atomic models identified as Complex-A (graphene/graphene/CNT, Complex-B (graphene/graphane/CNT), and Complex-C (graphane/graphane/ CNT). The two graphene-based nanoribbons have a length and width of 150 \r{A} and 21 \r{A}, respectively. The CNT used to trigger the self-scrolling mechanism is a conventional zigzag CNT $(0,22)$ with a diameter of 17.21 \r{A} and length of 100 \r{A}. The graphene nanoribbon and CNT have 1296 and 2112 carbon atoms, respectively. The initial distance between the nanoribbons and the CNT wall is 5.41 \r{A}.In the initial assembly, the nanoribbons are placed into contact with the nanotubes as shown in Figures \ref{fig:complexes}(d-f). The nanoribbons were placed on opposite sides regarding the CNT center.

The composite systems (Complexes --A, --B, and --C) were equilibrated/thermalized within an NVT ensemble at $300$, $600$, and $1000$ K using the Nos\'e-Hoover thermostat \cite{hoover1985canonical}. We considered 1000 K as the uppermost temperature limit, envisioned carbon-based optoelectronic applications. Scroll structures are resilient to high temperatures (up to 4000K), as recently demonstrated in references \cite{junior2021reactive,lee_Polymers}. The system equilibration was initially performed to eliminate any residual stress during 100 ps. The time step for the simulations was $0.1$ fs.  The MD snapshots and trajectories were obtained by using the free visualization and analysis program VMD \cite{humphrey1996vmd}.  

In the present work, we performed 10 MD runs (with different seeds) for each temperature. The random seeds defined distinct initial sets of atomic velocities in the equilibration of the systems. All the MD runs yielded the same products for each complex. In this sense, the results reported in the manuscript are the most representative outcomes obtained for all the simulations.

\section{Results and Discussions}

\begin{figure*}[h!]
\centering
\includegraphics[width=\linewidth]{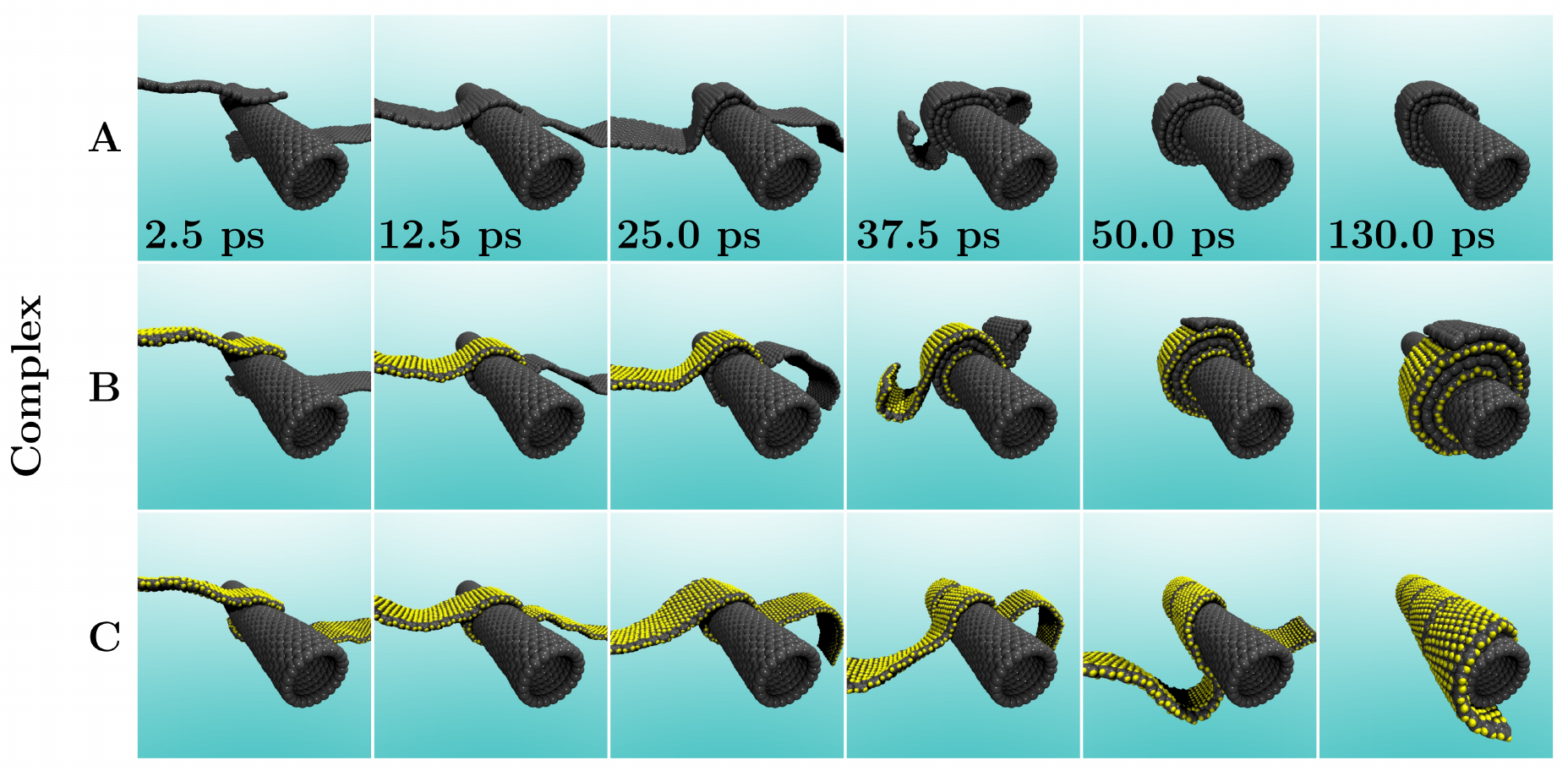}
\caption{Representative MD snapshots for the spontaneous self-scrolling process at 300 K, for the complexes presented in Figure \ref{fig:complexes}.}
\label{fig:MD}
\end{figure*}

We begin our discussion by presenting the dynamical behavior of the self-scrolling process for all complexes studied here (see Figure \ref{fig:complexes}). In Figure \ref{fig:MD} we show a sequence of representative MD snapshots (in lines) for Complex-A, -B, and -C from top to bottom, respectively, for simulations at $300$ K. In the very first stage of the simulation (about $2.5$ ps), the nanoribbons are in contact with the CNT surface, and the self-scrolling process takes place. This trend is realized in the simulations for all complexes, as shown in Figure \ref{fig:MD}. Such behavior also occurred for the self-scrolling mechanism of a single graphene nanoribbon \cite{xia2010fabrication}. 

At $12.5$ ps of simulation, the nanoribbons start to curl around the CNT. Subsequently, for Complexes A and B, the overlap between the nanoribbons occurs. The graphane nanoribbons in Complex-C do not form a scroll. Instead, it simply wraps the CNT by covering its surface and is temperature independent (for the temperature values considered here, up to $1000$ K), creating a lock-like mechanism (see Figure S1 in the Supplementary Material). This behavior is attributed to stereo hydrogen-hydrogen interactions when the graphane layers move to create an opposing force to the sliding. This trend is similar to observed for other hydrocarbon structures, such as the lander molecule, where this lock-like mechanism is present \cite{otero2004lock} and it is based on the same concept of the lock-and-key protein-protein interactions \cite{morrison2006lock}.It is worth mentioning that this process also occurred for the self-scrolling mechanism of a single graphene nanoribbon (see Supplementary Material).

Differences in the complexes dynamics become clear at $37.5$ ps. At this stage, in Complexes A and B, the nanoribbons are almost completely wrapped, as a consequence of the interlayer interaction presented by graphene and graphane \cite{savin2016symmetric,perez2015pi,savin_PRB}. Graphene/graphane bilayer (Complex-B) presents an interlayer interaction energy smaller than graphene/graphene one (Complex-A) \cite{umadevi_PCCP}. Even so, the self-scrolling process for both complexes takes the same time yielding a completely scrolled configuration at $50$ ps. For Complex-C, the graphane nanoribbons wrap the entire CNT wall at $130$ ps. In all cases, the nanoribbons showed discontinuous wrinkles and corrugations during the spontaneous self-scrolling process. Other reactive MD simulations performed within the scope of the AIREBO interatomic potential \cite{huang2015mechanical,song2013atomic} also pointed for such instabilities. The Supplementary Material includes the videos for the dynamics of the complexes A-C. It is worth mentioning that the degree of instability for the nanoribbons was more pronounced for temperatures higher than $300$ K.

The Supplementary Material presents the case in which a single graphane nanoribbon is interacting with a CNT. The main conclusion for the graphane/CNT interaction presented above holds, and the system evolves to a final state in which the graphane wraps the CNT wall. Isolated graphane nanoscrolls were already investigated by molecular dynamics simulations \cite{savin_PRB}, and the results are consistent with the ones presented here.

It is important to stress that depending on the area of the nanoribbon, the most stable configuration is not planar but a scrolled one. This trend is related to the size of the overlapped region that enables van der Waals gains to overcome the elastic cost to scroll the structures \cite{braga2004structure}. Even for small graphene nanobelts, the self-scrolling can be a self-sustained process \cite{martins_nanotechlogy}.

\begin{figure*}[h!]
\centering
\includegraphics[width=\linewidth]{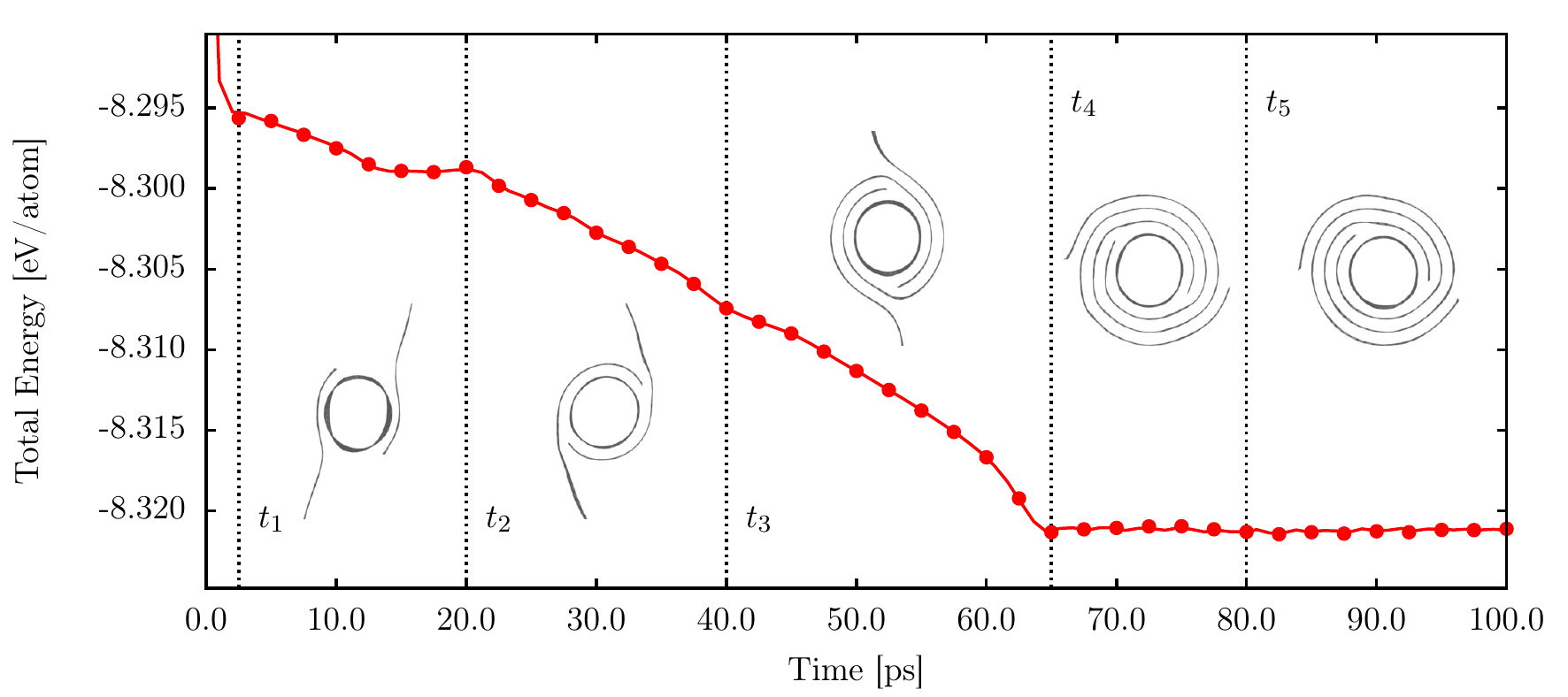}
\caption{Total energy time evolution as function of simulation time for the Complex-A at 10 K. The insets are MD snapshots extracted from the simulation trajectory at $t_1=2.5$, $t_2=20$, $t_3=40$, $t_4=65$, and $t_2=80$ ps, respectively.}
\label{fig:energy01}
\end{figure*} 

\begin{figure*}[h!]
\centering
\includegraphics[width=\linewidth]{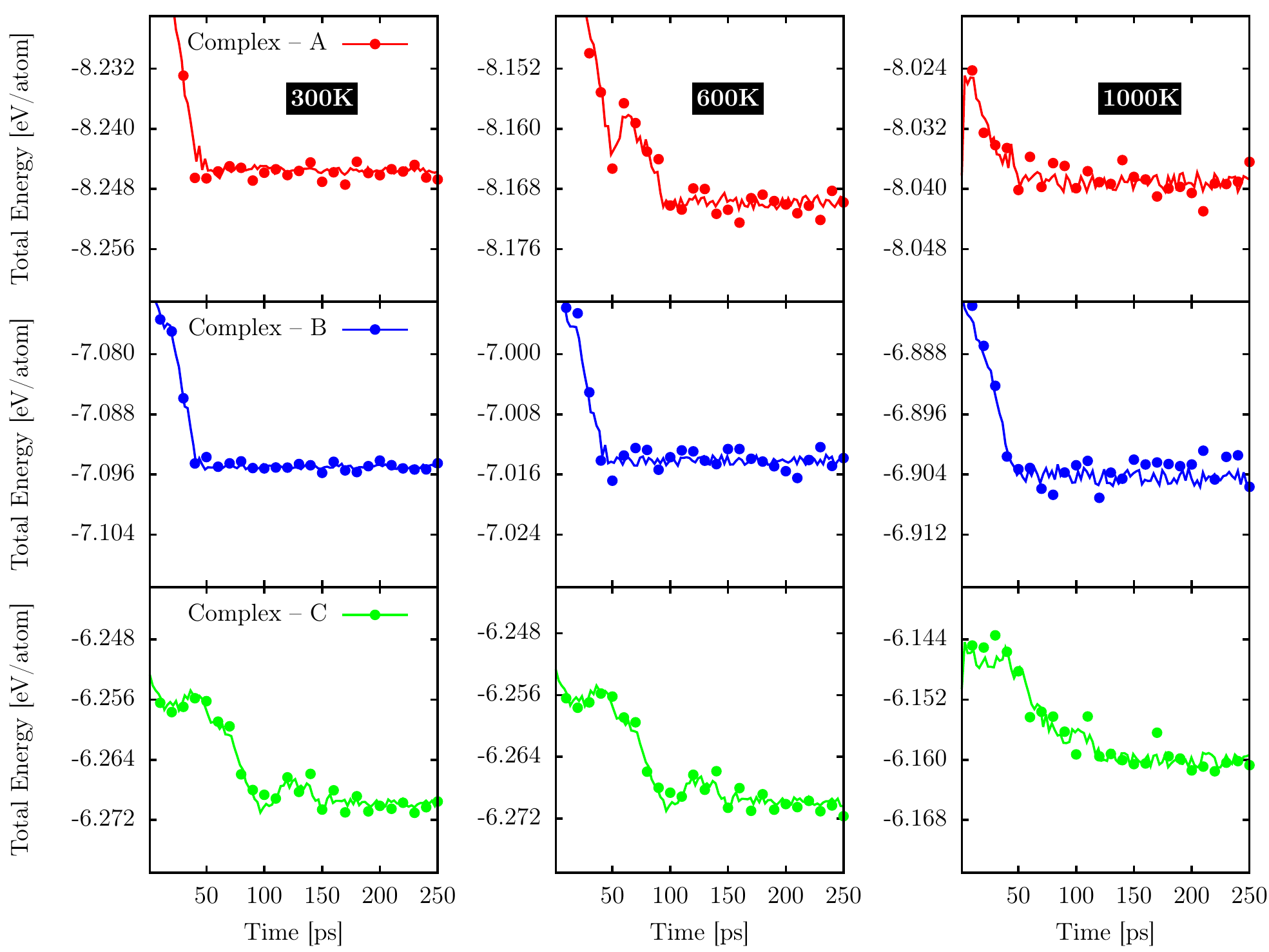}
\caption{Total energy time evolution as a function of the simulation time for all complexes at 300, 600, and 1000 K. The red, blue, and green data points refer to the total energy values during the simulation time for the Complex-A, Complex-B, and Complex-C, respectively. The lines were obtained from the data interpolation.}
\label{fig:energy02}
\end{figure*}

In Figure \ref{fig:energy01} we present the time evolution for the total energy for the Complex-A at $10$ K at five different stages. During the dynamical process for the CNSs formation, there is a linear downhill trend that ends at $65$ ps, which represents a completely formed scroll. From that moment on, the total energy stabilizes, and a flat region persists (neglecting thermal fluctuations, scroll ``breathing'' - see video) until the end of the simulation. For all the temperature regimes simulated here ($300$, $600$, and $1000$ K), the time evolution of the total energy shows a similar trend. An overall picture of the CNSs formation is presented in Figure \ref{fig:energy02}. The existence of an energy plateau is the signature of the scroll formation. At high temperatures, the CNSs total energies are larger due to the higher degree of atomic thermal vibrations, which tends to difficult the spontaneous scrolling process. Even so, one can realize a well-defined plateau in all the cases, standing for the scroll's formation. 

\section{Conclusions and remarks}

Through fully-atomistic reactive (ReaxFF) MD simulations, we have studied the hydrogenation effects on the self-scrolling process of graphene and graphane (hydrogenated graphene) nanoribbons interacting with a carbon nanotube (CNT). We have considered different configurations: (1) graphene/graphene/CNT; (2) graphene/graphane/CNT, (3) graphane/graphane/CNT; and different temperatures, $300$, $600$, and $1000$ K. Our results show that the scroll formation is possible for configurations (1) and (2), for all temperatures. Configuration (3) does not yield scrolls. The scroll formation is related to the existence of an energy plateau in the total energy as a function of the simulation time. The approach used here to assemble the nanoribbons and CNT is prototypical and extensible to other species of 2D materials beyond graphene. The usage of nanotubes for triggering the scrolling processes of sheets and nanoribbons was investigated for carbon and boron nitrides systems \cite{li_PCCP,perim_CPC,siahlo2018structure}. The tribology of carbon-based nanostructures has been an object of research in the last years \cite{guo2005coupled,lebedeva2009dissipation}. In this sense, the present study can stimulate further investigations along these lines. 

\section{Acknowledgements}

The authors gratefully acknowledge the financial support from Brazilian research agencies CNPq, FAPESP, and FAP-DF. M.L.P.J gratefully acknowledges the financial support from CAPES grant 88882.383674/2019-01. J.M.S. and D.S.G. thanks the Center for Computing in Engineering and Sciences at Unicamp for financial support through the FAPESP/CEPID Grants \#2013/08293-7 and \#2018/11352-7. J.M.S acknowledges CENAPAD-SP (Centro Nacional de Alto Desenpenho em São Paulo - Universidade Estadual de Campinas - UNICAMP ) for computational support process (proj842). L.A.R.J acknowledges the financial support from a Brazilian Research Council FAP-DF and CNPq grants $00193.0000248/2019-32$ and $302236/2018-0$, respectively. L.A.R.J and J.M.S acknowledge CENAPAD-SP for providing the computational facilities. L.A.R.J. gratefully acknowledges the financial support from IFD/UnB (Edital $01/2020$) grant $23106.090790/2020-86$. The authors acknowledge the National Laboratory for Scientific Computing (LNCC/MCTI, Brazil) for providing HPC resources of the SDumont supercomputer, which have contributed to the research results reported within this paper. URL: \url{http://sdumont.lncc.br}.

%\section{References}
\bibliographystyle{elsarticle-num}
\bibliography{cas-refs.bib}
\end{document}